\documentclass[
 reprint,
 amsmath,
 amssymb,
 superscriptaddress,
 aps,
 groupedaddress
]{revtex4-1}

\usepackage[utf8]{inputenc}
\usepackage{graphicx}
\usepackage{bm}
\usepackage{gensymb}
\usepackage{float}

\usepackage{array,mathtools,amssymb,booktabs}
\newcolumntype{C}{>{$}c<{$}}
\AtBeginDocument{
\heavyrulewidth=.08em
\lightrulewidth=.05em
\cmidrulewidth=.03em
\belowrulesep=.65ex
\belowbottomsep=0pt
\aboverulesep=.4ex
\abovetopsep=0pt
\cmidrulesep=\doublerulesep
\cmidrulekern=.5em
\defaultaddspace=.5em
}
\usepackage{bm}
\usepackage{float}
\usepackage{hyperref}
\usepackage{siunitx}
\begin{document}


\title{Measurements of electric quadrupole transition frequencies in $^{226}$Ra$^{+}$}
\author{C. A. Holliman}
\email{Corresponding author: cholliman3@gmail.com}
\author{M. Fan}
\author{A. M. Jayich}
\affiliation{Department of Physics, University of California, Santa Barbara, Santa Barbara, California 93106, USA}
\affiliation{California Institute for Quantum Entanglement, Santa Barbara, California 93106, USA}

\begin{abstract}
    We report the first driving of the $7s\ ^2S_{1/2}\rightarrow 6d\ ^2D_{3/2}$ and $7s\ ^2S_{1/2}\rightarrow 6d\ ^2D_{5/2}$ electric quadrupole (E2) transitions in Ra$^{+}$. We measure the frequencies of both E2 transitions, and two other low-lying transitions in $^{226}$Ra$^{+}$ that are important for controlling the radium ion's motional and internal states: $6d\ ^2D_{3/2}\rightarrow 7p\ ^2P_{3/2}^{o}$ and $6d\ ^2D_{5/2}\rightarrow 7p\ ^2P_{3/2}^{o}$.
\end{abstract}

\date{\today}

\maketitle
\section{Introduction}
The radium ion has two subhertz-linewidth electric quadrupole (E2) transitions from the ground state: $7s\ ^2S_{1/2}\rightarrow$ $6d\ ^2D_{5/2}$ at 728 nm and $7s\ ^2S_{1/2}\rightarrow$ $6d\ ^2D_{3/2}$ at 828 nm.  The E2 transition to the $D_{5/2}$ state, which has a lifetime of  $\sim$300 ms \cite{Pal2009}, is useful for electron shelving, ground-state cooling, and controlling an optical qubit, and it could also serve as the clock transition for an optical clock \cite{NunezPortela2014}. Measurement of these two quadrupole transitions across a chain of isotopes can be used to obtain information related to the nuclear structure of radium, such as the specific mass shift and the change in mean-square nuclear charge radii \cite{Heilig1974, reinhard2019charge}. The degree of nonlinearity in a King plot comparison of the two transitions could set bounds on new physics beyond the standard model \cite{Berengut2018}. There are 11 radium isotopes between mass number 213 and mass number 234 with half-lives longer than 1 min that could be ionized, trapped, and compared with a high precision on a King plot.  A precision King plot can also be made using one E2 transition in Ra$^+$ and the Ra $\ ^1S_0 \rightarrow \ ^3P_{0}$ intercombination line at 765 nm, which has a $2\pi\times5$ mHz linewidth \cite{Bieron2007}.

\begin{figure}[h]
    \centering
    \includegraphics[width=0.7\linewidth]{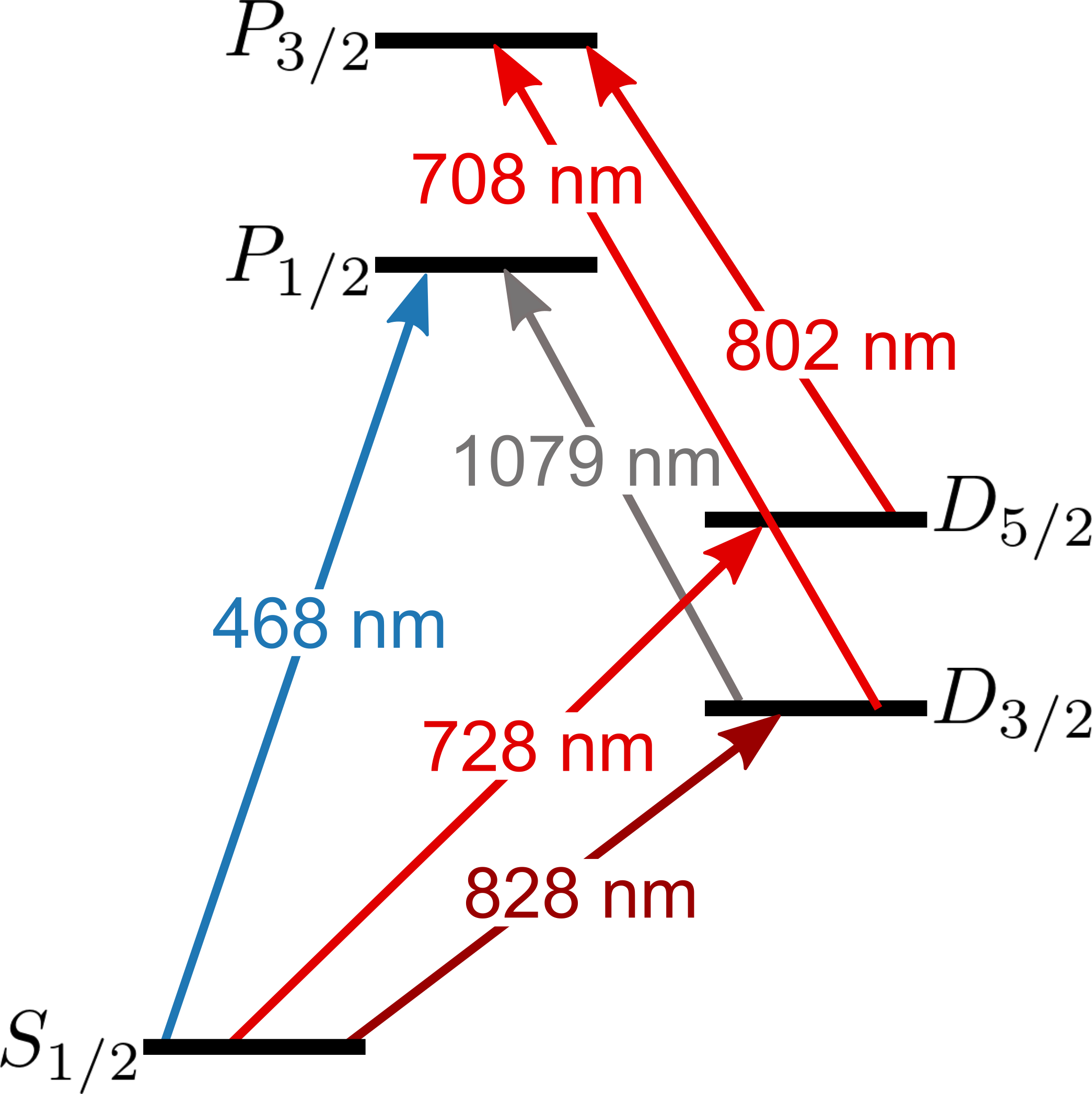}
    \caption{The Ra$^{+}$ energy level structure showing the transitions driven in this work.}  
    \label{fig:ra_energy_level}
\end{figure}

In this work we measure the two $7s\rightarrow 6d$ electric quadrupole transition frequencies, as well as the frequencies of the $6d\ ^2D_{3/2}\rightarrow$ $7p\ ^2P_{3/2}^{o}$ (708-nm) and $6d\ ^2D_{5/2}\rightarrow$ $7p\ ^2P_{3/2}^{o}$ (802-nm) electric dipole transitions (see Fig. \ref{fig:ra_energy_level}) in $^{226}$Ra$^+$ ($I=0$, 1600 yr half-life). Previously the only optical frequency measurement in $^{226}$Ra$^+$ with an uncertainty of less than \SI{4}{\giga\hertz} is that of the $7s\ ^2S_{1/2}\rightarrow$ $7p\ ^2P_{1/2}^{o}$ (468-nm) Doppler cooling transition \cite{Fan2019}. Combining that measurement with the measurements in this work, we calculate the $6d\ ^2D_{3/2}\rightarrow$ $7p\ ^2P_{1/2}^{o}$ (1079-nm) and $7s\ ^2S_{1/2}\rightarrow$ $7p\ ^2P_{3/2}^{o}$ (382-nm) frequencies, which are useful for laser-cooling Ra$^{+}$. 

For our measurements we use a single laser-cooled radium ion, which is loaded by ablating a 10 $\mu$Ci RaCl$_{2}$ target $\sim$\SI{15}{\milli\meter} from the trap. The radio-frequency trapping voltage is turned on \SI{20}{\micro\second} after ablation to enhance the loading efficiency. The radium ion trap and loading procedure used in this work have been described previously \cite{Fan2019}.  We use a heated iodine vapor cell as a frequency reference. The Ra$^+$ spectroscopy transitions and the iodine reference are driven with a tunable Ti:sapphire laser, whose frequency is recorded with a wave meter (High Finesse WS-8) to determine the Ra$^+$ transition frequencies from the known iodine reference lines.  We describe the Ra$^+$ transition measurements and fits in Sec. \ref{sec:radium-spectroscopy}, and the iodine frequency reference spectroscopy in Sec. \ref{sec:iodine-spectroscopy}. From the combined iodine and Ra$^+$ data we determine the Ra$^+$ transition frequencies in Sec. \ref{sec:frequencies}, and with these values we present an updated King plot that includes radium-226 in Sec. \ref{sec:king}. 

\section{Radium Spectroscopy}\label{sec:radium-spectroscopy}

The Ra$^{+}$ transitions are measured using state detection, where the ion is bright if the population is in the cooling cycle and dark otherwise.  Bright-state fluorescence photons at 468 nm are collected onto a photomultiplier tube and the counts are then time-tagged with respect to the measurement pulse sequences \cite{Pruttivarasin2015}. In 1 ms of state detection 35 photons are collected on average if the population is in the $S_{1/2}$ or $D_{3/2}$ ``bright'' state, and 1.5 photons of background scattered light if the population is shelved in the $D_{5/2}$ ``dark'' state. We set the bright-state detection threshold to 12 counts.

All Ra$^+$ spectroscopy pulse sequences begin with 0.5 ms of laser cooling and an initial state detection (see SD1 in Fig. \ref{fig:ps728}).  If the state detection finds the ion in the dark state, then the data point is excluded because the ion is not properly initialized.  All pulse sequences finish with a second state detection step and then optical pumping to remove any remaining population from the $D_{5/2}$ state. We give a detailed description of the $7s\ ^2S_{1/2}\rightarrow 6d\ ^2D_{5/2}$ electric quadrupole ``clock'' transition measurement.  The other measurements are similar, with brief descriptions provided.

\vspace{-12pt}
\subsection{$7s\ ^2S_{1/2}\rightarrow 6d\ ^2D_{5/2}$ (728 nm)}
\begin{figure}[h]
    \centering
    \includegraphics[width=1.\linewidth]{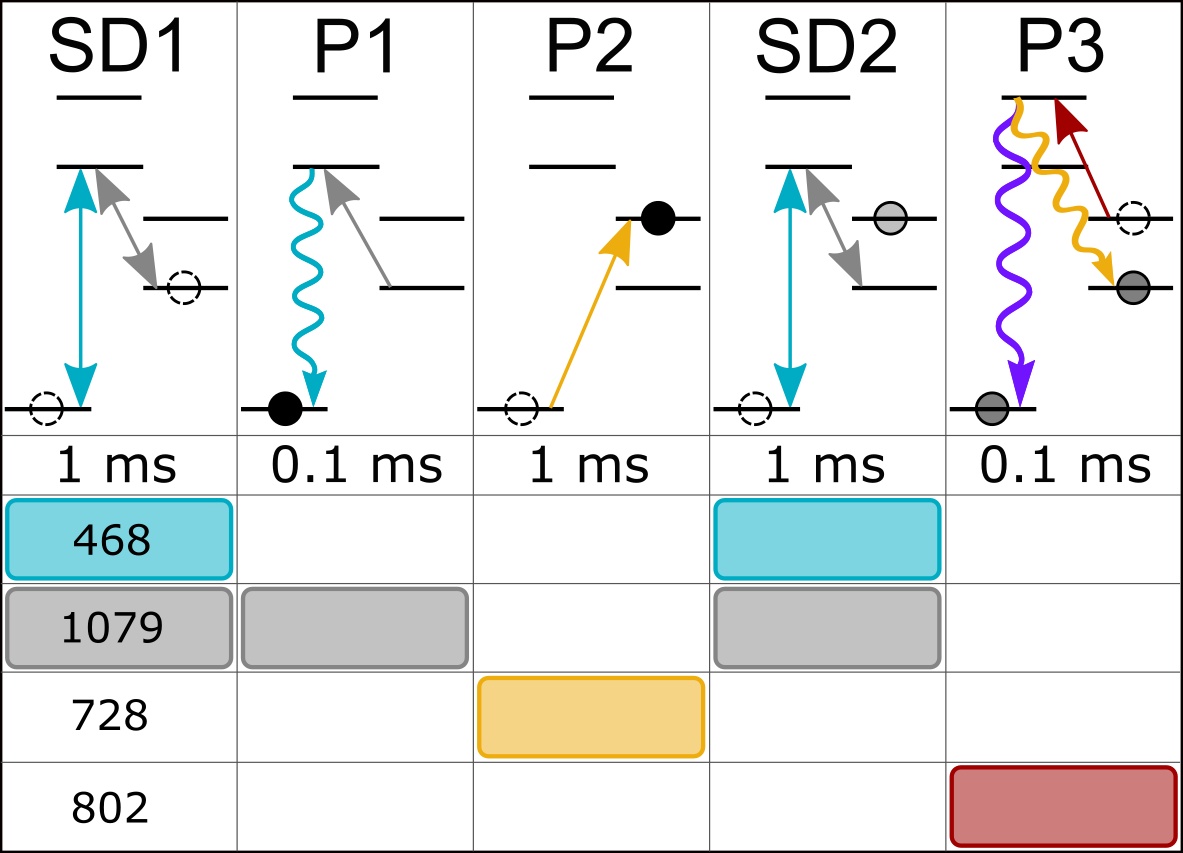}
    \caption{The pulse sequence for the $^2S_{1/2}\rightarrow ^2D_{5/2}$ E2 transition measurement. Squiggly lines depict E1 allowed decays, straight lines show optical pumping transitions, and double-arrows indicate optical cycling transitions.}  
    \label{fig:ps728}
\end{figure}

\begin{figure}[h!]
    \centering
    \includegraphics[width=0.95\linewidth]{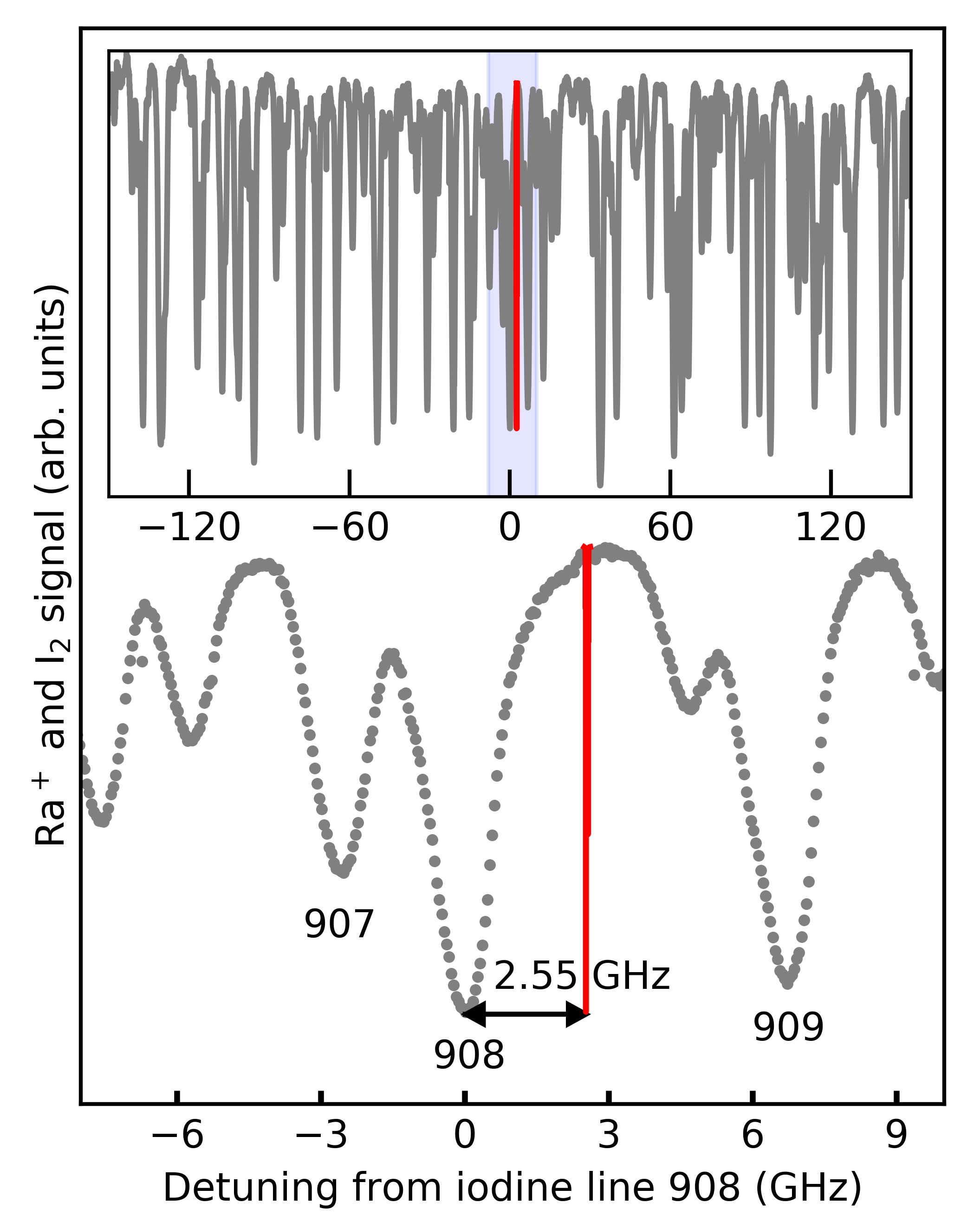}
    \caption{The iodine absorption spectrum is plotted with the Ra$^{+}$ scan, where the iodine data (gray) and Ra$^{+}$ data (red; see Fig. \ref{fig:polarizations}) are scaled and offset to highlight the detuning between the transitions. We use the line indices from \cite{Gerstenkorn1982,Gerstenkorn1982a} to denote iodine lines. Inset: Iodine lines in a 300-GHz range around the closest reference line, 908, with the outset region highlighted in blue.}  
    \label{fig:joint_plot}
\end{figure}

\vspace{-12pt}
The pulse sequence for the clock transition measurement is shown in Fig. \ref{fig:ps728}.  Before each measurement we Doppler-cool the ion for 0.5 ms. The initial state detection  determines whether the ion is cooled, and whether the population is in a bright state (SD1).  Any population in the $D_{3/2}$ state is then optically pumped with light at 1079 nm to the ground state (P1). The $7s\ ^2S_{1/2}\rightarrow 6d\ ^2D_{5/2}$ spectroscopy transition is then driven with light at 728 nm for 1 ms (P2), and if the light is on resonance, the population can be shelved to the $D_{5/2}$ state, and the shelving probability is measured with a second state detection (SD2). To prepare for the next measurement any shelved population is cleaned out with light at 802 nm, which drives the population to the $S_{1/2}$ and $D_{3/2}$ bright states (P3).  Over many pulse sequences the 728-nm laser is swept over $\sim$\SI{100}{\mega\hertz} to drive all possible transitions between Zeeman levels of the two states. The $7s\ ^2S_{1/2}\rightarrow 6d\ ^2D_{5/2}$ spectroscopy is shown in Fig. \ref{fig:joint_plot}, along with an inset of the corresponding iodine absorption reference spectrum. 

\begin{figure*}[ht]
    \centering
    \includegraphics[width=\textwidth]{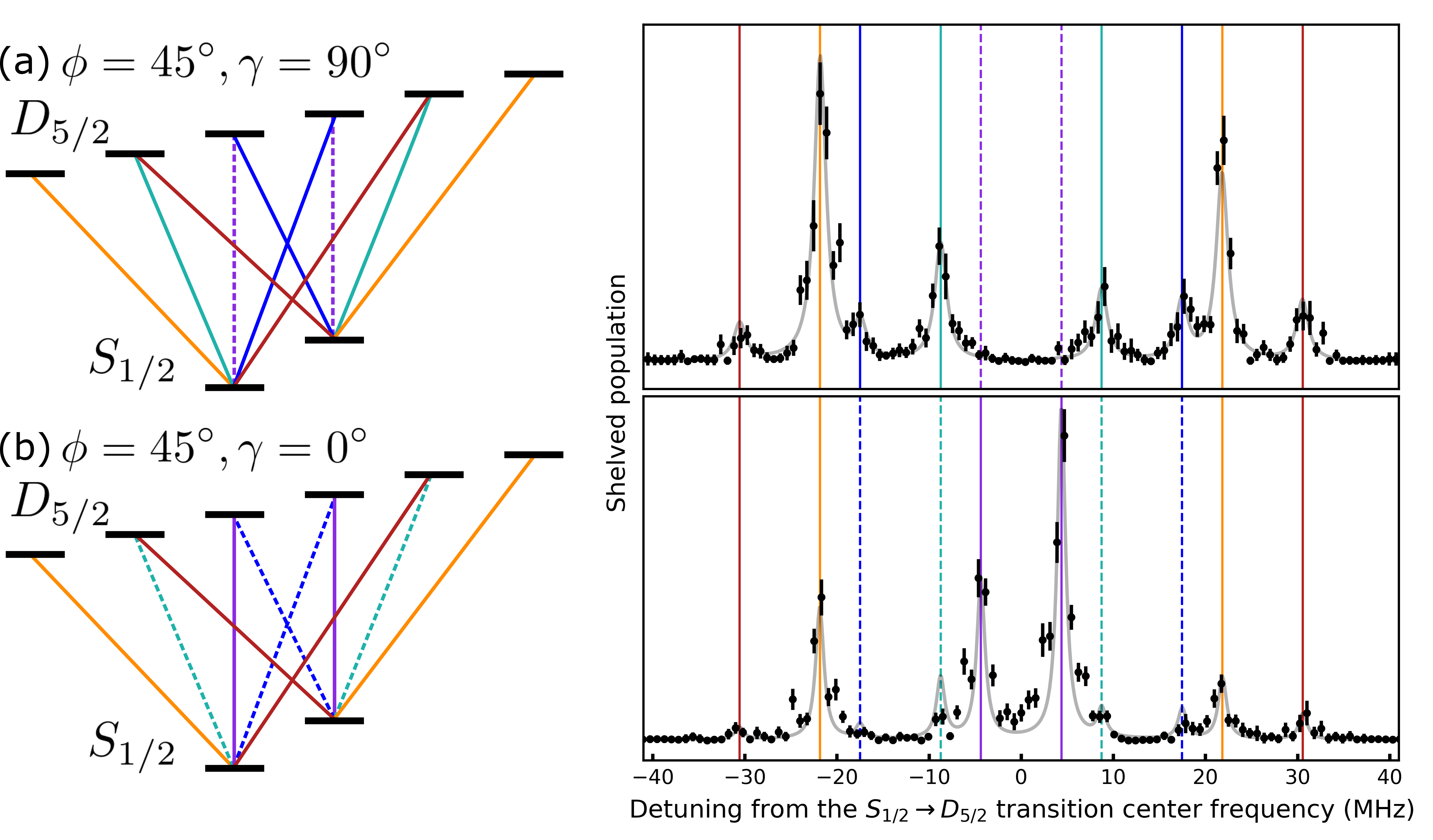}
    \caption{Two measurements of the $S_{1/2}\rightarrow D_{5/2}$ transition with angle $\phi=45^\circ$ between the $k$-vector and the magnetic field: (a) angle $\gamma=90^\circ$ between the laser polarization and the magnetic field, which suppresses $\Delta m= 0$; (b) $\gamma=0^\circ$, which suppresses $\Delta m =\pm 1$. Dashed lines indicate transitions suppressed by the choice of $\phi$ and $\gamma$.  Symmetric pairs of transitions are the same color. Error bars are the most likely 68\% confidence interval of a binomial distribution.}
    \label{fig:polarizations}
\end{figure*}

To resolve the $S_{1/2}\rightarrow D_{5/2}$ and $S_{1/2}\rightarrow D_{3/2}$ Zeeman sub-structure of the transitions, a 7.8-G magnetic field is applied parallel to the trap's radio-frequency rods, which we define as the $z$ axis. The magnetic field spreads the $S_{1/2}\rightarrow D_{5/2}$ transitions across $\sim$\SI{60}{\mega\hertz}. The energy splitting due to the applied magnetic field is calculated from $\Delta E_{j} = g_{j} \mu_{\text{B}} B  m$, where $g_{j}$ is the Land\'e $g$-factor for state with total angular momentum $j$, $\mu_{\text{B}}$ is the Bohr magneton, $B$ is the magnetic field, and $m$ is the magnetic quantum number. The Rabi frequency for transitions between the $S_{1/2,m}$ and the $D_{5/2,m'}$ Zeeman levels is

\begin{equation}
\begin{aligned}
    \Omega = &\bigg|\frac{e E_{0}}{2\hbar} \langle S_{1/2}||r^{2}\mathbf{C}^{(2)}||D_{5/2}\rangle \\
    & \sum_{j=-2}^{2} \begin{pmatrix}
    1/2 & 2 & 5/2 \\
    -m & \Delta m & m'
    \end{pmatrix} g^{(\Delta m)}\bigg|,
\end{aligned}
\end{equation}

\noindent where $\langle S_{1/2}||r^{2}\mathbf{C}^{(2)}||D_{5/2}\rangle$ is the reduced matrix element, the summation is over Wigner 3-$j$ symbols and a geometry-dependent factor, $g^{(\Delta m)}$ \cite{James1998, Roos2000}, given by

\begin{equation}
\begin{aligned}
   g^{(0)} = &\frac{1}{2}\Big|\cos(\gamma)\sin(2\phi)\Big|,\\
   g^{(\pm 1)} = &\frac{1}{\sqrt{6}}\Big|\cos(\gamma)\cos(2\phi) + i\sin(\gamma)\cos(\phi)\Big|,\\
   g^{(\pm 2)} = &\frac{1}{\sqrt{6}}\Big|\frac{1}{2}\cos(\gamma)\sin(2\phi) + i\sin(\gamma)\sin(\phi)\Big|,
\end{aligned}
\end{equation}

\noindent where $\Delta m$ is the change in magnetic quantum number, $\phi$ is the angle between the laser's $k$ vector and the magnetic field, and $\gamma$ is the angle between the laser polarization and the magnetic field vector projected into the plane of incidence.

For E2 transitions $\Delta m = 0, \pm 1, \pm 2$ are allowed, which gives rise to 10 transitions between Zeeman levels for $S_{1/2}\rightarrow$ $D_{5/2}$. We measure these transitions using $\phi=45^\circ$, and two values of $\gamma$, $0^\circ$ and $90^\circ$, which suppress certain $|\Delta m|$ transitions (see Fig. \ref{fig:polarizations}). We fit the spectroscopy data to a sum of 10 Lorentzians, from which we extract the applied magnetic field, $\phi$, $\gamma$, and the $S_{1/2}\rightarrow$ $D_{5/2}$ transition frequency. Due to the 468- and 1079-nm laser polarizations, during Doppler cooling the ground-state $m = -1/2$ level is preferentially populated, and so reduces the probability of transition occurring from the $m=1/2$ level.  The ground state population imbalance is one of the fitting parameters. The $\sim$0.5-Hz natural linewidth of the $S_{1/2}\rightarrow$ $D_{5/2}$ transition is broadened by laser power and magnetic field fluctuations. There are small micromotion sidebands in the spectrum at our trap drive frequency, \SI{1.8}{\mega\hertz}. We average the two center frequencies extracted from the two fits of the spectra at different values of $\gamma$ to determine the $S_{1/2}\rightarrow D_{5/2}$ transition frequency. 

\begin{figure}[h]
    \centering
    \includegraphics[width=0.9\linewidth]{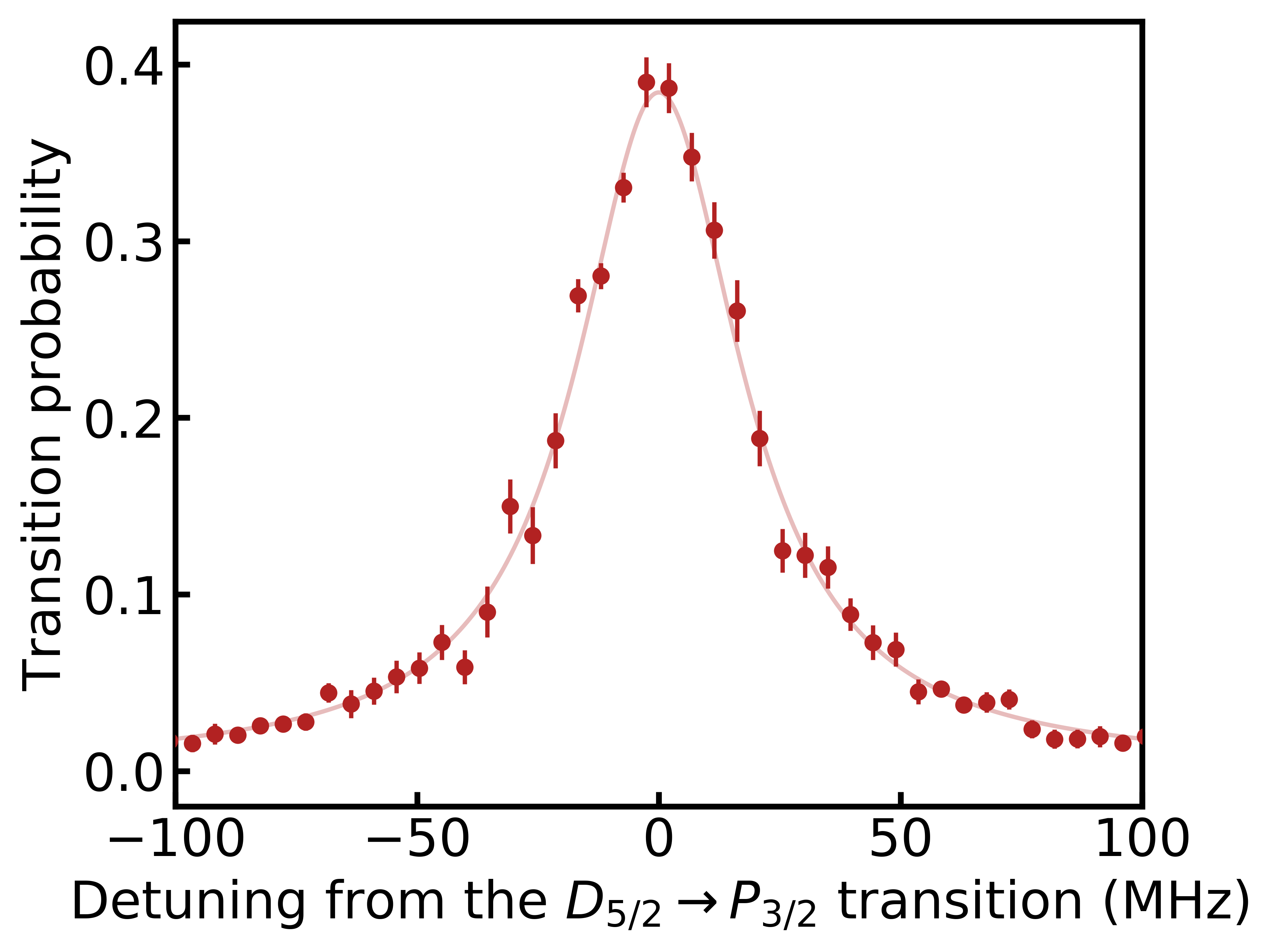}
    \caption{Spectroscopy of the $D_{5/2}\rightarrow$ $P_{3/2}$ transition. A Lorentzian fit gives an FWHM of \SI{41.5\pm0.9}{\mega\hertz}.}  
    \label{fig:802}
\end{figure}

\vspace{-12pt}
\subsection{$6d\ ^2D_{5/2}\rightarrow 7p\ ^2P_{3/2}^{o}$ (802 nm)}
\vspace{-12pt}
For measuring both the $^2D_{5/2}\rightarrow\ ^2P_{3/2}$ and the $^2D_{3/2} \rightarrow\ ^2P_{3/2}$ transitions a magnetic field of $\sim$2.5 G is applied to prevent coherent dark states \cite{Berkeland2002}.  After initialization, the ion is optically pumped to the $D_{5/2}$ state with light at 468 and 708 nm for \SI{60}{\micro\second}. Any population in the $D_{5/2}$ state is then pumped with the spectroscopy light at 802 nm for \SI{150}{\micro\second}. If the light is on resonance, the population can be driven to the $P_{3/2}$ state, where decays will populate the bright states (see Fig. \ref{fig:802}).  After the second state detection the $D_{5/2}$ state is optically pumped with resonant 802-nm light from an external cavity diode laser. 

To determine the line center for the $D_{5/2}\rightarrow P_{3/2}$ transition we fit the data to a Lorentzian (see Fig. \ref{fig:802}).  The full width at half-maximum (FWHM) of \SI{41.5\pm0.9}{\mega\hertz} gives a lower bound of \SI{3.84\pm0.09}{\nano\second} for the $P_{3/2}$-state lifetime, which agrees with the calculated value of \SI{4.73}{\nano\second} \cite{Pal2009}. The 802-nm laser light is incident on the ion along the trap's \textit{z} axis to minimize micromotion broadening \cite{Berkeland1998}.  We measured the 802-nm laser's beam waist to be \SI{300\pm50}{\micro\meter} and our beam power to be \SI{1.2\pm0.1}{\micro\W}. We estimate that the transition is power broadened by $\sim$\SI{8}{\mega\hertz} from the 802-nm beam waist and power.

\vspace{-12pt}
\subsection{$7s\ ^2S_{1/2}\rightarrow 6d\ ^2D_{3/2}$ (828 nm)}
\vspace{-12pt}
After initialization, any population in the $D_{3/2}$ state is optically pumped with light at 1079 nm for \SI{100}{\micro\second} to the ground state. The $7s\ ^2S_{1/2}\rightarrow 6d\ ^2D_{3/2}$ transition is driven with light at 828 nm for 10 ms, and if the light is on resonance, the ion can be shelved to the $D_{3/2}$ state. Any population in the $D_{3/2}$ state is then pumped with light at 708 nm for \SI{60}{\micro\second}, which drives the population to the $P_{3/2}$ state whose decays populate the $S_{1/2}$ and $D_{5/2}$ states. 

The population is shelved to the $D_{5/2}$ state with a low fidelity due to the branching fraction from the $P_{3/2}$ state.  The shelved probability is determined from a binomial distribution of $k$ shelved events (corresponding to fewer than 12 photons collected during the second state detection) in $n$ trials (where the ion is properly initialized at the beginning of the pulse sequence).  There are eight transitions between Zeeman levels for the $S_{1/2}\rightarrow$ $D_{3/2}$ transition. We fit the spectra in a similar fashion as the $S_{1/2}\rightarrow$ $D_{5/2}$ transition to extract the center frequency.

\vspace{-12pt}
\subsection{$6d\ ^2D_{3/2}\rightarrow 7p\ ^2P_{3/2}^{o}$ (708 nm)}
\vspace{-12pt}
After initialization, any population in the $S_{1/2}$ state is optically pumped with light at 468 nm for \SI{100}{\micro\second} to the $D_{3/2}$ state. The population is then pumped with light at 708 nm for \SI{50}{\micro\second}, and if the light is on resonance, the population can be driven to the $P_{3/2}$ state whose decays populate the $S_{1/2}$ and $D_{5/2}$ bright states. The data are fit to a Lorentzian to determine the center frequency.

\section{Iodine Frequency Reference}\label{sec:iodine-spectroscopy}

For iodine absorption spectroscopy the Ti:sapphire laser is scanned at 10 MHz per second across a frequency range that includes at least two iodine reference lines, as well as the target Ra$^+$ transition.  To reduce drift in the wave meter measurement we measure the iodine spectra both before and after Ra$^{+}$ spectroscopy. We use one iodine line as the frequency reference, and we determine the line's frequency by fitting the line's absorption dip in both scans to a Voigt function and averaging the two centers.  We then calibrate the iodine line frequency we measured with a wave meter to an absolute frequency using IodineSpec5 \cite{IodineSpec5, Knoeckel2004}, which provides a comprehensive I$_{2}$ reference data set based on iodine spectrum measurements including the original iodine atlas work \cite{Gerstenkorn1982, *Gerstenkorn1982a}.  We fit the corresponding absorption dip in the IodineSpec5 data set to a Voigt function to determine the absolute frequency of the measured iodine line center.

The $^{127}$I$_{2}$ vapor cell frequency reference (75 mm long, with a 19-mm outer diameter and windows angled at 11$^\circ$) is heated in a tube furnace to  $\sim$500$\degree$C, the temperature at which the iodine atlas lines were measured \cite{Gerstenkorn1982, *Gerstenkorn1982a}.  We scan the temperature of the I$_{2}$ cell by $\pm50\degree$C around 500$\degree$C and vary an applied magnetic field by $\pm4$ G, and for both we find frequency shifts within the fitting uncertainty. To compensate for laser power drifts the power is recorded on a reference photodiode before the iodine cell.

\section{Ra Transition Frequencies}\label{sec:frequencies}

The Ra$^{+}$ transition frequencies are calculated from the frequency difference between the radium and the iodine reference line centers that are recorded with a wave-meter.  The fit iodine reference lines calculated in IodineSpec5 are used to calibrate the wave-meter frequencies.

The closest iodine reference line to the $S_{1/2}\rightarrow D_{5/2}$ transition is line 908 from \cite{Gerstenkorn1982, Gerstenkorn1982a}, which we calculate using IodineSpec5 to be \SI{412004.754\pm0.015}{\giga\hertz}.  From the difference between radium and iodine spectroscopy center fits, \SI{2.947}{\giga\hertz}, we determine a transition frequency of \SI{412007.701\pm0.018}{\giga\hertz}.  We take the total uncertainty in our measurements to be $\sigma_{\text{total}} = \sqrt{\sigma_{\text{Ra}^{+}}^{2} + \sigma_{\text{I}_{2}}^2 + \sigma_{\text{spec}}^{2} + \sigma_{\text{wm}}^{2}}$,  where $\sigma_{\text{Ra}^{+}}$ is the radium fitting uncertainty, $\sigma_{\text{I}_{2}}$ is the measured iodine fitting uncertainty, $\sigma_{\text{spec}}$ is the IodineSpec5 line uncertainty, and $\sigma_{\text{wm}}$ is the wavemeter uncertainty, 10 MHz. The IodineSpec5 line uncertainties range from 2 to 45 MHz for the lines referenced in this work.  The IodineSpec5 fitting uncertainties are on the kilohertz level and are negligible compared to the other uncertainties. 

\begin{table}[h] \centering
\caption{Summary of $^{226}$Ra$^{+}$ frequency measurements. All units are GHz. *Frequencies calculated from measurements. $^{\dagger}$Frequencies extrapolated from a King plot \cite{Giri2011a}.
\label{table:frequency_summary}}
\begin{ruledtabular}
\begin{tabular}{lcccc}
    Transition & \cite{Rasmussen1933} & \cite{NunezPortela2014} & This work, \cite{Fan2019}  \\ 
    \midrule
    $S_{1/2}\rightarrow$ $P_{3/2}$ & \SI{785723\pm4}{} & \SI{785721.670\pm0.070}{}* & \SI{785722.10\pm0.03}{}* \\
    $S_{1/2}\rightarrow$ $P_{1/2}$ & \SI{640092\pm7}{} & \SI{640096.647\pm0.023}{}* & \SI{640096.63\pm0.06}{} \cite{Fan2019} \\
    $D_{3/2}\rightarrow$ $P_{3/2}$ & \SI{423438\pm6}{} & $\cdots$ & \SI{423444.39\pm0.03}{} \\
    $S_{1/2}\rightarrow$ $D_{5/2}$ & $\cdots$ & $\cdots$ & \SI{412007.701\pm0.018}{} \\
    $D_{5/2}\rightarrow$ $P_{3/2}$ & \SI{373717\pm5}{} & $\cdots$ & \SI{373714.40\pm0.02}{} \\
    $S_{1/2}\rightarrow$ $D_{3/2}$ & $\cdots$ & \SI{362277.361\pm0.033}{}$^{\dagger}$ & \SI{362277.68\pm0.05}{} \\
    $D_{3/2}\rightarrow$ $P_{1/2}$ & $\cdots$ & \SI{277819.285\pm0.018}{}$^{\dagger}$ & \SI{277818.95\pm0.08}{}* \\
\end{tabular}
\vspace*{-\baselineskip}
\end{ruledtabular}
\end{table}

The reported value for the $D_{3/2}\rightarrow P_{1/2}$ transition, \SI{277818.95\pm0.08}{\giga\hertz}, comes from the frequency difference between the $S_{1/2} \rightarrow D_{3/2}$ E2 transition measured in this work and the measurement of the $S_{1/2} \rightarrow P_{1/2}$ transition by Fan, \emph{et al.} \cite{Fan2019}. 

From our measurements we can calculate the $S_{1/2}\rightarrow P_{3/2}$ (382-nm) transition frequency by summing either the 728- and 802-nm frequencies, \SI{785722.10\pm0.03}{\giga\hertz}, or the 828- and 708-nm frequencies, \SI{785722.07\pm0.05}{\giga\hertz}.  Both pairs of transitions originate in the ground state and end in the $P_{3/2}$ state.  The two calculated frequencies are in good agreement, and we report the value calculated from the 728- and 802-nm frequencies, as the uncertainty in the underlying measurements is smaller.  The measured and calculated frequencies are summarized in Table \ref{table:frequency_summary}.

There are discrepancies between the transition frequencies measured in this work and those reported by Nunez, \textit{et al.} \cite{NunezPortela2014} (see Table \ref{table:frequency_summary}).  We are able to resolve some of the discrepancies by redoing the analysis in \cite{NunezPortela2014}.  The $D_{3/2}\rightarrow$ $P_{1/2}$ transition frequency reported in \cite{NunezPortela2014}, \SI{277819.285\pm0.018}{\giga\hertz}, is extrapolated from a King plot with data from Giri \textit{et al.} \cite{Giri2011a}. From the 1079 nm/482 nm King plot (Fig. 4 in \cite{Giri2011a}) we find a slope of \SI{-0.342\pm0.007}{} and a $y$-intercept of \SI{-2.2\pm0.5}{\tera\hertz} amu, which gives the transformed isotope shift [see Eq. (4) below] of 54.5(1.3)\SI{}{\tera\hertz} amu and the $D_{3/2}\rightarrow$ $P_{1/2}$ transition frequency in $^{226}$Ra$^+$ of \SI{277819.2\pm0.3}{\giga\hertz}. This value agrees with our measurement. The $S_{1/2}\rightarrow$ $D_{3/2}$ transition frequency reported in \cite{NunezPortela2014} is calculated from the frequency difference between the $S_{1/2}\rightarrow$ $P_{1/2}$ and the $D_{3/2}\rightarrow$ $P_{1/2}$ transitions. With the recalculated value of the  $D_{3/2}\rightarrow$ $P_{1/2}$ frequency, the $S_{1/2}\rightarrow$ $D_{3/2}$ transition frequency is \SI{362277.4\pm0.3}{\giga\hertz}, which is also in agreement with our measurement. There remains the discrepancy between the $S_{1/2}\rightarrow$ $P_{3/2}$ frequency measured in this work and the value reported in \cite{NunezPortela2014}. This discrepancy could be resolved by direct spectroscopy of the S$_{1/2}$-to-$P_{3/2}$ transition in one of many radium isotopes, including isotope(s) 212, 214, or 221-226 \cite{Neu1988}.

\section{King Plot}\label{sec:king}

\begin{figure}[h]
    \centering
    \includegraphics[width=1.0\linewidth]{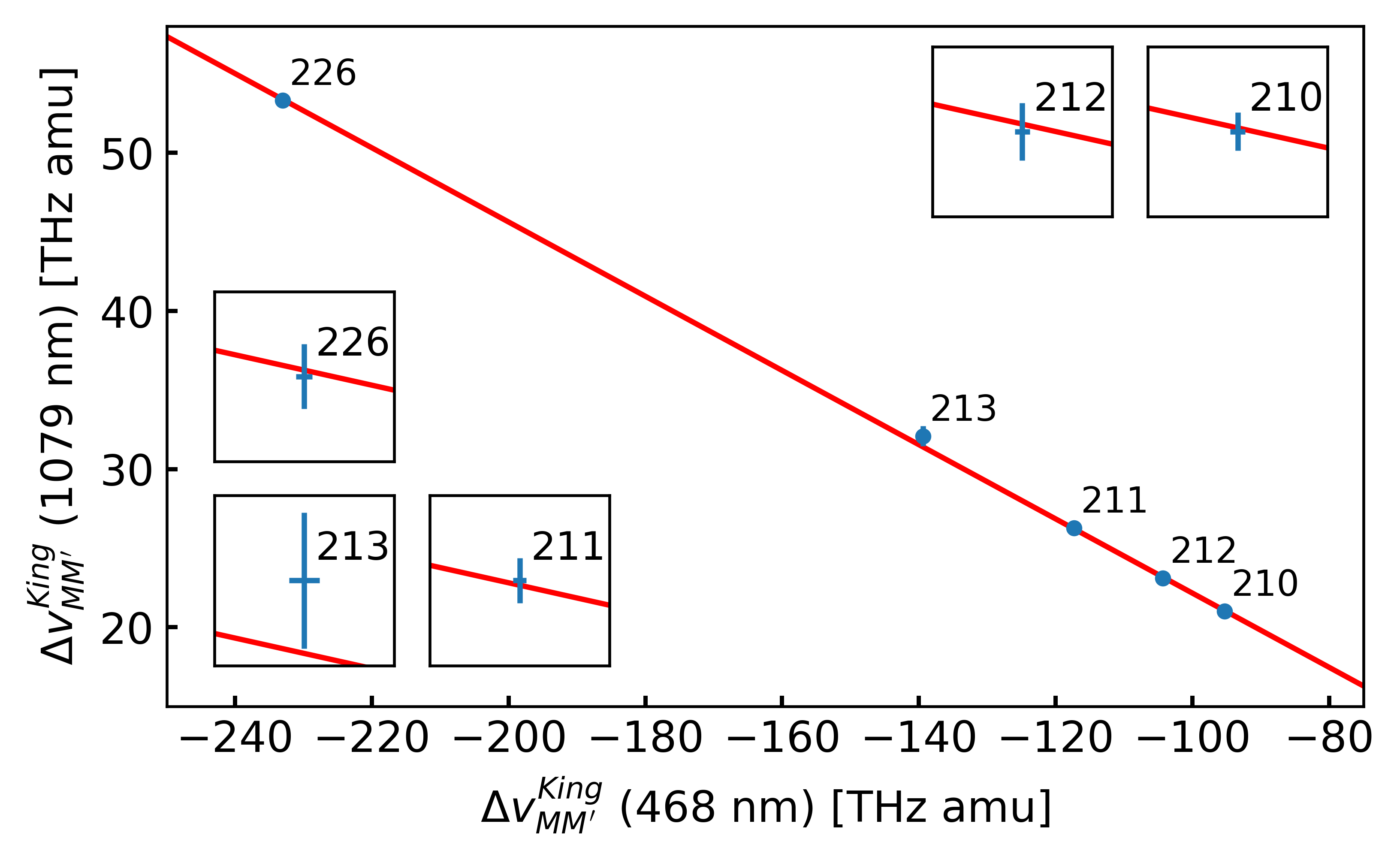}
    \caption{A King plot of the transformed isotope shifts of the $D_{3/2}\rightarrow$ $P_{1/2}$ versus the  $S_{1/2}\rightarrow$ $P_{1/2}$ transitions. The reference isotope is $^{214}$Ra$^{+}$. Insets: Magnifications of x200 for the 468 nm axis and x50 for the 1079 nm axis.}
    \label{fig:king}
\end{figure}

We determine isotope shifts for the 708- and 1079-nm transitions in $^{226}$Ra$^{+}$. The isotope shift between a target and a reference isotope is $\delta\nu_{MM'} = \nu_{M} - \nu_{M'}$, where $M$ is the target isotope nuclear mass, $M'$ is the reference isotope nuclear mass, and $\nu$ is the transition frequency. We use $^{214}$Ra$^{+}$, which has a closed neutron shell, as the reference isotope \cite{Wendt1987}.  The 1079-nm $^{214}$Ra$^{+}$ reference frequency is \SI{277805.656}{\giga\hertz} \cite{Giri2011a} and the 708-nm $^{214}$Ra$^{+}$ reference frequency is \SI{423434.989}{\giga\hertz} \cite{Versolato2010}. Isotope shifts of 468, 708, and 1079 nm are listed in Table \ref{table:isotope_shifts}. The isotope shift is parameterized as 
\begin{equation}
        \delta\nu_{MM'}= (K_{\text{NMS}}+K_{\text{SMS}})\frac{M-M'}{MM'}+F_{\text{FS}}\lambda_{MM'},\\
\end{equation}
where $K_{\text{NMS}}$ and $K_{\text{SMS}}$ are the normal and specific mass shifts, $F_{\text{FS}}$ is the field shift, and $\lambda_{MM'}$ is the Seltzer moment, which to lowest order is the difference in mean square nuclear charge radii, $\lambda_{MM'} \approx  \delta\langle r^{2}\rangle_{MM'}$ \cite{Heilig1974}. The transformed isotope shift is
\begin{equation} \label{shift}
\begin{aligned}
    \Delta \nu_{M M'}^{\text{King}} = \delta\nu_{M M'}\frac{MM'}{M-M'}-K_{\text{NMS}},\\
\end{aligned}
\end{equation}

\noindent from which the ratio of field shifts between transitions and the difference in specific mass shifts can be determined \cite{King1963}.  King plot comparisons of the transformed isotopes shifts give the Ra$^+$ field shift ratios, $F_{708}/F_{468}=-0.171(3)$ and $F_{1079}/F_{468}=-0.235(2)$, which are summarized in  Table \ref{table:field_shift}. The 1079 nm/468 nm King plot is shown in Fig. \ref{fig:king}.

\begin{table}[H] \centering
\caption{Measured Ra$^+$ isotope shifts of the 468-, 708-, and 1079-nm transitions for $^{210-213}$Ra$^+$ and $^{226}$Ra$^+$ relative to $^{214}$Ra$^+$. All units are MHz. The 468 nm isotope shifts are from \cite{Wendt1987}. The 708- and 1079-nm isotope shifts for $^{210-213}$Ra$^+$ are from \cite{Wansbeek2012}. *An isotope shift calculated in this work.
\label{table:isotope_shifts}}
\begin{ruledtabular}
\begin{tabular}{lcccc}
    Isotope & 468 nm & 708 nm & 1079 nm \\ 
    \midrule
    210 & 8449(6) & $\cdots$ &  -1884(16)  \\
    211 & 7770(4) & $\cdots$ & -1755(14) \\
    212 & 4583(3) & -701(20) & -1025(12)  \\
    213 & 3049(3) & -453(34) & -707(14) \\
    214 & 0 & 0 &  0 \\
    226 & \SI{-57852\pm18}{} & \SI{9403\pm27}{}* & \SI{13294\pm76}{}* \\
\end{tabular}
\vspace{2pt}
\vspace*{-\baselineskip}
\end{ruledtabular}
\end{table}

\begin{table}[h] \centering
\caption{$^{226}$Ra$^{+}$ field shift ratios.
\label{table:field_shift}}
\begin{ruledtabular}
\begin{tabular}{lcccc}
    FS Ratio & Theory \cite{Wansbeek2012} & Experiment \cite{Wansbeek2012} & This work \\ 
    \midrule
    $F_{708}/F_{468}$ & -0.16(1) & -0.136(57) & -0.171(3) \\
    $F_{1079}/F_{468}$ & -0.23(2) & -0.244(11) & -0.235(2) \\
\end{tabular}
\end{ruledtabular}
\end{table}

\section{Conclusion}
The first driving of the 728- and 828-nm electric quadrupole transitions in Ra$^{+}$ lays the groundwork for quantum information science and precision measurement experiments with Ra$^+$.  Isotope shift spectroscopy of Ra$^+$ dipole transitions has been done at the ISOLDE facility at CERN, and could be extended to a high precision by trapping Ra$^+$ and measuring the E2 transitions \cite{Lynch2018}. Estimates for the nonlinearities of a King plot of radium's two narrow linewidth transitions following \cite{Flambaum2018} could guide searches to constrain new physics. 

We thank A. Vutha and R. Ruiz for feedback on the manuscript and acknowledge support from NSF Grant No. PHY-1912665 and the Office of the President, University of California (Grant No. MRP-19-601445).

\end{document}